\documentclass[ conference, letterpaper, 10pt, times]{IEEEtran}
\IEEEoverridecommandlockouts
\usepackage{amsmath,amssymb,bm}
\usepackage{graphicx}
\usepackage{booktabs}
\usepackage{array}
\usepackage{url}
\usepackage{algorithm}
\usepackage{algpseudocode}
\newcommand{\tr}{\operatorname{tr}}
\newcommand{\Tset}{\mathcal{T}}
\newcommand{\Pset}{\mathcal{P}}

\newcommand{\zcap}{\zeta_{\mathrm{cap}}}

\usepackage{cite}
\usepackage[T5]{fontenc}
\usepackage[utf8]{inputenc}
\usepackage{amsmath,amssymb,amsfonts}
\usepackage{textcomp}
\usepackage{xcolor}
\usepackage{tabularx}
\usepackage{stfloats}
\usepackage{multirow}
\usepackage[caption=false,font=footnotesize]{subfig}     
\usepackage{url}
\DeclareRobustCommand*{\IEEEauthorrefmark}[1]{%
  \raisebox{0pt}[0pt][0pt]{\textsuperscript{\footnotesize #1}}%
  }
\def\BibTeX{{\rm B\kern-.05em{\sc i\kern-.025em b}\kern-.08em
    T\kern-.1667em\lower.7ex\hbox{E}\kern-.125emX}}
    
\IEEEaftertitletext{\vspace{-2.5\baselineskip}} 
\begin{document}

\title{Conflict-Aware Impedance-Sensitivity-Based Tuning for Multi-Mode Damping Enhancement}

\author{
    \IEEEauthorblockN{
        Quang Manh Hoang\IEEEauthorrefmark{1},
        Xuan Thanh Trinh\IEEEauthorrefmark{1},
        Sina Zarrabian\IEEEauthorrefmark{3},
        Akhtar Hussain\IEEEauthorrefmark{2},
        Van-Hai Bui\IEEEauthorrefmark{1,*}
    }

    \IEEEauthorblockA{
        \IEEEauthorrefmark{1}
        Department of Electrical and Computer Engineering,
        University of Michigan-Dearborn, USA
    }

    \IEEEauthorblockA{
        \IEEEauthorrefmark{2}
        Department of Electrical and Computer Engineering,
        Laval University, Canada
    }

    \IEEEauthorblockA{
        \IEEEauthorrefmark{3}
        Department of Electrical and Computer Engineering,
        United States Naval Academy, Annapolis, MD, USA
    }

    \thanks{
        \textit{Corresponding author:
        V.H. Bui (vhbui@umich.edu)}
    }

    \thanks{
        \textit{*DISCLAIMER: The views expressed in this
        article are those of the author(s) and do not
        reflect the official policy or position of the
        U.S. Naval Academy, Department of the Navy,
        the Department of War, or the U.S. Government.}
    }
}

\maketitle
\setlength{\abovedisplayskip}{6pt plus 2pt minus 2pt}
\setlength{\belowdisplayskip}{6pt plus 2pt minus 2pt}
\setlength{\abovedisplayshortskip}{0pt plus 2pt}
\setlength{\belowdisplayshortskip}{4pt plus 2pt minus 2pt}

\begin{abstract}
Impedance-based eigenvalue sensitivity provides a practical way to relate converter parameters to poorly damped system modes, but tuning becomes nontrivial when one parameter improves one target mode while degrading another. This paper introduces a conflict-aware tuning strategy for simultaneous damping enhancement of multiple modes. The method converts impedance-derived eigenvalue sensitivities into normalized damping-ratio sensitivities, diagnoses opposing modal tuning directions, and selects either one parameter or a coordinated pair at each iteration. A low-dimensional max--min problem maximizes the predicted damping of the worst target mode subject to parameter limits, protected-mode constraints, and implementation-specific feasibility requirements. Each accepted update is verified on the full system model and the sensitivities are recomputed. Studies on a four-bus system and a modified IEEE 14-bus system demonstrate the value of pairwise coordination, including a case in which single-parameter tuning stops before the predefined damping target is reached.
\end{abstract}


\section{Introduction}
Power systems with a high penetration of inverter-based resources (IBRs) can exhibit several poorly damped oscillatory modes whose locations depend strongly on converter controls, network conditions, and interactions among multiple devices \cite{Wang2019}. Classical state-space modal analysis provides direct access to eigenvalues, participation factors, and parameter sensitivities, but it requires complete internal models of the system, which may be unavailable when equipment from different vendors is represented by grey- or black-box models\cite{hoang2026fedavgim}. Impedance-based methods are therefore attractive because they can assess small-signal stability from terminal characteristics \cite{Cespedes2014} and network interconnections \cite{Rygg2016} without requiring disclosure of all internal controller states \cite{Sun2011}.

In the classical source--load formulation, impedance-based stability analysis requires the system to be partitioned into two subsystems at a selected interface \cite{Sun2011}. This can be difficult in large multi-converter networks, where several devices may interact across the system. Stability-scanning methods reduce the dependence on a single predefined partition by evaluating the system at different locations, but the number of required scans grows with system size and operating conditions. More importantly, conventional impedance-based screening mainly indicates whether an undesired interaction exists and where it occurs, while providing limited information on which device or control parameter influences a poorly damped mode or how that parameter should be adjusted. Recent work has extended impedance analysis toward modal participation and parameter sensitivity. In particular, impedance participation analysis relates whole-system residues to modal participation \cite{Zhu2022Participation}, while the eigenvalue-sensitivity framework in \cite{Zhu2023Sensitivity} quantifies how a system pole changes with an apparatus parameter. These developments provide the directional information needed for damping-oriented retuning.

Parameter sensitivities and coordinated control have also been employed for oscillation damping enhancement \cite{Harnefors2007}. A centralized multimode power oscillation damping strategy was developed to damp multiple oscillatory modes through coordinated converter control \cite{jankovic2023centralised}. Eigenvalue sensitivities have been applied to the design of active- and reactive-power damping controllers for grid-forming converters \cite{mateu2025power}. Optimization-based approaches have further coordinated damping and inertia parameters across multiple converter-interfaced resources under small-signal stability constraints \cite{feng2025hybrid}. 

However, a fundamental challenge arises when the same parameter improves one target mode while degrading another, leading to conflicting tuning directions. Repeated single-parameter updates may therefore alternate between modes, degrade an already improved mode, or reach a point where no admissible single-parameter change can further improve the weakest target mode. Existing coordinated tuning methods can optimize several parameters, but they do not explicitly use conflicts among modal sensitivities, together with parameter limits, to determine when single-parameter tuning becomes insufficient and coordinated changes are required. In addition, many existing approaches rely on white-box state-space models \cite{jankovic2023centralised,mateu2025power,feng2025hybrid}, which may not be available in converter-dominated multi-vendor systems.

To address these issues, this paper proposes a conflict-aware impedance-sensitivity-based tuning framework for multi-mode damping enhancement. The method identifies poorly damped modes from the whole-system impedance model \cite{Zhu2022Participation}, converts impedance-derived eigenvalue sensitivities into normalized damping-ratio sensitivities, and detects conflicting tuning directions. A single-parameter update is used when a suitable direction exists, otherwise, a pair of parameters is coordinated through a low-dimensional max--min problem to improve the weakest target mode while satisfying parameter and protected-mode constraints. Each accepted update is verified on the full system, after which the target modes are tracked and the sensitivities are recomputed. The impedance-based formulation makes the framework applicable to multi-vendor systems where detailed controller models may be proprietary or unavailable.

The method is evaluated on a four-bus system and a modified IEEE 14-bus system. The four-bus case illustrates how coordinated parameter changes can compensate for conflicting modal effects, while the IEEE 14-bus case shows that pairwise coordination can reach the prescribed damping target when single-parameter tuning stalls. An equal-budget comparison based on an $\ell_1$ limit on the normalized parameter change is also included to distinguish the benefit of coordination from that of allowing a larger total parameter change.

\section{Impedance-Based Damping Sensitivity and Tuning Directions}

\subsection{Impedance-Based Damping Sensitivity}

Using impedance models provided by different vendors, a whole-system impedance model can be constructed and the system modes can be identified \cite{Zhu2022Participation,Zhu2023Sensitivity}. The objective of this subsection is to convert the impedance-based eigenvalue sensitivity into a normalized damping-ratio sensitivity that can be directly used for parameter tuning.

Consider the eigenvalue associated with mode $m$,
\begin{equation}
    \lambda_m = \sigma_m + j\omega_m,
\end{equation}
where $\sigma_m<0$ denotes the decay rate and $\omega_m$ is the damped angular frequency. The corresponding damping ratio is
\begin{equation}
    \zeta_m =
    -\frac{\sigma_m}{\sqrt{\sigma_m^2+\omega_m^2}}.
    \label{eq:zeta}
\end{equation}

For a tunable parameter $p_j$ associated with apparatus $k$, the impedance-based eigenvalue sensitivity is given by \cite{Zhu2023Sensitivity}
\begin{equation}
    g_{mj}
    = \frac{\partial \lambda_m}{\partial p_j}
    = -\tr\!\left[
    \frac{\partial Z_k(\lambda_m)}{\partial p_j}R_{km}
    \right],
    \label{eq:eig_sens}
\end{equation}
where $Z_k$ is the apparatus impedance representation and $R_{km}$ is the residue block associated with mode $m$ in the whole-system impedance/admittance formulation. Equation~\eqref{eq:eig_sens} describes the local movement of mode $m$ caused by a small variation in parameter $p_j$. Since the sensitivity is evaluated at the current parameter point, it provides a first-order local description rather than a prediction for an arbitrary finite parameter change.

The eigenvalue sensitivity in \eqref{eq:eig_sens} describes movement in the complex plane, whereas the tuning objective in this work is expressed directly in terms of modal damping ratios. In addition, the candidate parameters may have different physical units and nominal magnitudes. To place all tuning variables on a common scale, define the normalized parameter deviation as
\begin{equation}
    x_j = \frac{p_j-p_{j0}}{p_{j0}},
    \label{eq:norm}
\end{equation}
where $p_{j0}$ is the nominal value. Thus, $\Delta x_j=0.05$ represents a $5\%$ change from the nominal value independent of the physical unit of $p_j$.

Differentiating \eqref{eq:zeta} and applying the chain rule gives the normalized damping-ratio sensitivity
\begin{equation}
H_{mj}
= \frac{\partial \zeta_m}{\partial x_j}
= p_{j0}
\frac{-\omega_m^2\operatorname{Re}(g_{mj})
+\sigma_m\omega_m\operatorname{Im}(g_{mj})}
{(\sigma_m^2+\omega_m^2)^{3/2}}.
\label{eq:H}
\end{equation}
The quantity $H_{mj}$ directly relates a normalized change in parameter $p_j$ to the resulting local change in the damping ratio of mode $m$. Its sign indicates the favorable tuning direction, while its magnitude indicates how strongly the mode responds to the parameter variation.

For a sufficiently small normalized parameter update $\Delta\bm{x}$, the damping ratios can then be approximated by the first-order model
\begin{equation}
    \bm{\zeta}^{\mathrm{pred}}
    = \bm{\zeta} + H\Delta\bm{x},
    \label{eq:pred}
\end{equation}
where $H$ collects the normalized damping sensitivities of the monitored modes to the available parameters. This linear model provides a common basis for comparing parameter effects across multiple modes and is used in the next subsection to identify compatible and conflicting tuning directions.

\subsection{Compatible and Conflicting Tuning Directions}

Let $\Tset$ denote the target modes whose damping ratios must be improved. For two target modes $a,b\in\Tset$, the local effect of parameter $j$ is classified as
\begin{equation}
H_{aj}H_{bj}
\begin{cases}
>0, & \text{compatible},\\
<0, & \text{conflicting}.
\end{cases}
\label{eq:direction_class}
\end{equation}

For a compatible parameter, the same tuning direction improves both modes: the parameter is increased when both sensitivities are positive and decreased when both are negative. Such parameters are preferred and can be adjusted until the damping requirement is reached or the allowable parameter limit is encountered. If the available compatible directions reach their limits before all target modes satisfy the damping requirement, conflicting parameters must also be considered.

For a conflicting parameter, either tuning direction improves one mode while degrading the other. Such a parameter is therefore not preferred for an independent update, but it may still be useful when coordinated with another parameter whose effect compensates the damping reduction. A pair of parameter directions can thus provide a combined update that improves the weakest target mode while preventing unacceptable degradation of the others.

Because a sign conflict may be insignificant when one sensitivity is very small, sensitivity magnitudes are also considered when selecting candidate parameters. This classification provides the basis for the single- and pairwise tuning strategy introduced in the next section.

\section{Conflict-Aware Impedance-Sensitivity-Based Tuning and Verification}

This section presents the proposed conflict-aware tuning framework, which uses the damping sensitivities developed in Section~II to select and verify single- or two-parameter updates for improving the weakest target mode.

At each iteration, the active parameter set $J$ satisfies
\begin{equation}
    1\le |J|\le 2,
    \qquad
    \Delta x_j=0,\quad \forall j\notin J.
    \label{eq:sparse}
\end{equation}
For each singleton or pair $J$, the local tuning problem is formulated as
\begin{subequations}
\label{eq:maxmin}
\begin{align}
\max_{\Delta\bm{x}_J}\quad
& \min\left\{
\zcap,\;
\min_{m\in\Tset}
\left(
\zeta_m+H_{m,J}\Delta\bm{x}_J
\right)
\right\}
\label{eq:maxmin_obj}\\
\text{s.t.}\quad
& \bm{x}_{\min}
\le \bm{x}+\Delta\bm{x}
\le \bm{x}_{\max},
\label{eq:maxmin_box}\\
& \|\Delta\bm{x}\|_{\infty}
\le \Delta x_{\max},
\label{eq:maxmin_step}\\
& \zeta_r+H_{r,J}\Delta\bm{x}_J
\ge \underline{\zeta}_r,
\quad \forall r\in\Pset,
\label{eq:maxmin_protected}\\
& c_q(\bm{x}+\Delta\bm{x})\le0,
\quad q=1,\ldots,n_c.
\label{eq:maxmin_impl}
\end{align}
\end{subequations}

The objective in \eqref{eq:maxmin_obj} maximizes the predicted damping ratio of the weakest target mode, with $\zcap$ limiting unnecessary improvement once the target damping level is reached. The constraints enforce the admissible parameter range, the maximum normalized step size, the minimum damping requirement of protected non-target modes, and implementation-specific feasibility conditions.

Since at most two parameters are active, each candidate problem is only one- or two-dimensional and can be solved directly over its feasible region. To minimize the number of parameters adjusted, the best single-parameter update is chosen if it provides at least a fraction $\eta\in(0,1]$ of the damping improvement predicted by the best two-parameter update. Otherwise, the two-parameter update is selected. The preference threshold $\eta$ is user-defined.

Because \eqref{eq:pred} is a local approximation, the selected update is verified on the full system before acceptance. The complete procedure is summarized in Algorithm~\ref{alg:conflict_tuning}.

\begin{algorithm}[t]
\caption{Conflict-Aware Sparse Max--Min Tuning}
\label{alg:conflict_tuning}
\begin{algorithmic}[1]
\Require $\bm{x}$, $\Tset$, $\Pset$, parameter bounds, $\zeta_{\mathrm{req}}$, $\eta$
\While{$\min_{m\in\Tset}\zeta_m < \zeta_{\mathrm{req}}$}
    \State Build the system model and compute the sensitivity matrix $H$.
    \State Form admissible singleton and pairwise sets $J$.
    \State Solve \eqref{eq:maxmin_obj}--\eqref{eq:maxmin_impl} for each $J$.
    \State Select the best singleton or pair using the $\eta$ threshold.
    \State Set $\alpha=1$.
    \While{the update is not accepted}
        \State Evaluate $\Delta\bm{x}_{\mathrm{trial}}
        =\alpha\Delta\bm{x}$ on the full system.
        \If{all constraints are satisfied and \eqref{eq:accept} holds}
            \State Accept the update:
            $\bm{x}\leftarrow\bm{x}+\Delta\bm{x}_{\mathrm{trial}}$.
        \Else
            \State $\alpha\leftarrow\alpha/2$.
        \EndIf
        \If{no admissible trial step remains}
            \State \Return no improving step found.
        \EndIf
    \EndWhile
    \State Recompute the modes and sensitivities at the new parameter point.
\EndWhile
\State \Return $\bm{x}$
\end{algorithmic}
\end{algorithm}

For each trial, the relevant modal branches and the complete eigenvalue spectrum are checked. A step is accepted only if all parameter and implementation constraints remain feasible, the protected modes satisfy their damping limits, no unacceptable weak or unstable mode appears, and
\begin{equation}
    \min_{m\in\Tset}\zeta_m^{\mathrm{new}}
    >
    \min_{m\in\Tset}\zeta_m^{\mathrm{current}}
    +\epsilon,
    \label{eq:accept}
\end{equation}
with improvement tolerance $\epsilon>0$. Thus, an individual target mode may decrease slightly as long as the weakest target improves and all protection constraints remain satisfied. If no trial step passes the full-system verification, the algorithm terminates within the current parameter bounds and two-parameter search space.

Finally, \eqref{eq:maxmin_step} allows each active parameter to change by up to $\Delta x_{\max}$, so a pair may have a larger total normalized change than a singleton. To separate this effect from the benefit of coordination, an equal-budget comparison additionally imposes
\begin{equation}
    \|\Delta\bm{x}\|_1\le B
    \label{eq:l1}
\end{equation}
on both singleton and pairwise candidates, where $B>0$ is the shared total-step budget. The pair then retains its additional tuning direction without receiving a larger total normalized-change budget.

The following case studies evaluate whether this framework can improve multiple poorly damped modes and resolve cases in which single-parameter tuning becomes insufficient.

\section{Case Studies}
\label{sec:case-studies}

The proposed method is evaluated on a four-bus system with a
grid-forming (GFM) converter at bus~3 and a grid-following (GFL)
converter at bus~4, and on a modified IEEE 14-bus system with
tunable GFM converters at buses~2, 3, and~8. Two poorly damped
modes are targeted in each case, with a required damping ratio of
$10\%$. Table~\ref{tab:setup} summarizes the main test settings.

The tunable GFM parameters are
$\mathcal P_{\rm GFM}=\{X_{\rm ov},D_{\rm w},f_{\rm droop},
f_{{\rm v},dq},f_{{\rm i},dq}\}$, while the GFL parameters are
$\mathcal P_{\rm GFL}=\{f_{{\rm v},dc},f_{\rm PLL},
f_{\tau,\rm PLL},f_{{\rm i},dq}\}$. All listed parameters are
eligible for tuning.

\begin{table}[t]
\caption{Test-System and Tuning Setup}
\label{tab:setup}
\centering
\footnotesize
\setlength{\tabcolsep}{3pt}
\renewcommand{\arraystretch}{1.10}
\begin{tabular}{@{}p{0.33\columnwidth}
                  p{0.27\columnwidth}
                  p{0.27\columnwidth}@{}}
\toprule
Item & Four-bus & IEEE 14-bus\\
\midrule
Target frequencies (Hz)
& 14.29, 27.35
& 10.36, 17.94\\

Initial damping (\%)
& 6.54, 8.79
& 2.5, 2.51\\

Tunable parameters
& 9 (GFM, GFL)
& 15 (GFM)\\

Bounds $p_j/p_{j0}$
& $[0.8,1.2]$
& $[0.6,1.4]$\\

Protected branches
& 8; $10\%$ floors
& 14; $10\%$ floors\\
\bottomrule
\end{tabular}
\end{table}

For both systems, $\Delta x_{\max}=0.05$, $\eta=0.90$, and $\epsilon=10^{-7}$. The implementation constraints include $f_{vdq}\ge1.5f_{idq}$. Modal tracking uses
MAC$\ge0.9$, and the maximum number of accepted updates is 60.
Four tuning rules are compared: a weakest-mode single-parameter
rule, a multi-mode singleton baseline, the proposed single/pair
rule, and an equal-total-step variant of the proposed method with
$B=0.05$. All methods use the same parameter
bounds, modal protection, tracking, and full-spectrum checks.

\begin{table}[t]
\caption{Final Target Damping Ratios and Accepted Updates}
\label{tab:results}
\centering
\footnotesize
\setlength{\tabcolsep}{2pt}
\begin{tabular}{@{}lrrrrrr@{}}
\toprule
& \multicolumn{3}{c}{Four-bus system} & \multicolumn{3}{c}{IEEE 14-bus system}\\
\cmidrule(lr){2-4}\cmidrule(l){5-7}
Rule & $\zeta_1$ (\%) & $\zeta_2$ (\%) & Updates
     & $\zeta_1$ (\%) & $\zeta_2$ (\%) & Updates\\
\midrule
Weakest-mode single & 9.86 & 9.70 & 19 & 4.45 & 2.98 & 12\\
Multi-mode single & 10.04 & 10.02 & 20 & 6.24 & 6.24 & 15\\
Single/pair & 10.15 & 10.04 & 10 & 10.02 & 10.02 & 32\\
Single/pair, equal step & 10.02 & 10.02 & 17 & 10.01 & 10.01 & 46\\
\bottomrule
\end{tabular}
\end{table}

\subsection{Four-Bus Case: Coordinating Conflicting Effects}

\begin{figure}[t]
\centering
\includegraphics[scale=0.85]
{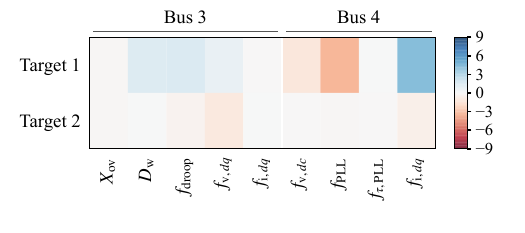}
\caption{Nominal four-bus damping sensitivities. Colors show
$20H_{mj}$, corresponding to the predicted damping change for
a $+20\%$ parameter variation.}
\label{fig:fourbus-sensitivity}
\end{figure}

Figure~\ref{fig:fourbus-sensitivity} shows several conflicting
parameter effects. For example, a $20\%$ increase in the bus-4
current-loop bandwidth produces predicted damping changes of
$+3.8$ and $-0.49$ percentage points for targets~1 and~2,
respectively. Significant conflicts also appear in the bus-3
droop-filter and voltage-loop bandwidths. In contrast, reducing
the bus-4 PLL bandwidth benefits both targets.

\begin{figure}[t]
\centering
\includegraphics[scale=0.9]
{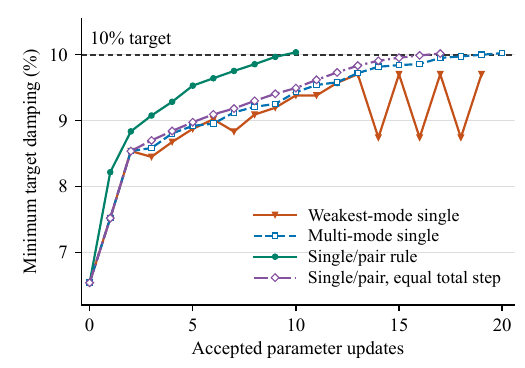}
\caption{Four-bus minimum target damping versus accepted updates.
The dashed line denotes the $10\%$ requirement.}
\label{fig:fourbus-damping}

\end{figure}

Figure~\ref{fig:fourbus-damping} compares the tuning rules.
The weakest-mode rule alternates between the two targets and
stops after 19 accepted updates at $9.7\%$. The multi-mode
singleton reaches the $10\%$ requirement in 20 updates, whereas
the proposed single/pair method requires only 10, with final
damping ratios of $10.15\%$ and $10.04\%$. Under the
equal-total-step constraint, the proposed method still reaches
the target in 17 updates, showing that its benefit is not solely
due to a larger total parameter movement.

\begin{figure}[t]
\centering
\includegraphics[scale=0.9]
{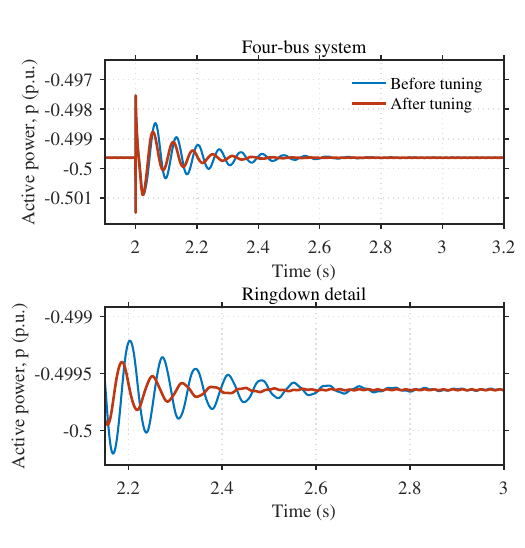}
\caption{Four-bus EMT active-power response at the bus-4 GFL
converter before and after tuning.}
\label{fig:fourbus-emt}
\end{figure}

The EMT results in Fig.~\ref{fig:fourbus-emt} provide
time-domain confirmation of the damping improvement. The tuned
case exhibits a faster-decaying oscillatory response, with the
peak-to-peak power variation over $2.4$--$3.0$~s decreasing from
$2.4\times10^{-4}$ to $4.5\times10^{-5}$~p.u., corresponding
to an approximately $81\%$ reduction.

\subsection{IEEE 14-Bus Case: Overcoming the Limitations of Singleton Tuning}

\begin{figure}[t]
\centering
\includegraphics[scale=0.85]
{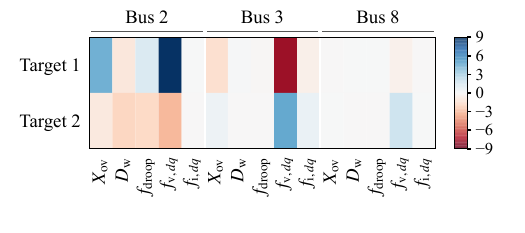}
\caption{Nominal IEEE 14-bus damping sensitivities, using
the same $20H_{mj}$ definition and scale as
Fig.~\ref{fig:fourbus-sensitivity}.}
\label{fig:ieee14-sensitivity}

\end{figure}

Figure~\ref{fig:ieee14-sensitivity} shows strong conflicting
effects, particularly in the voltage-loop bandwidths at
buses~2 and~3. A $+20\%$ perturbation predicts damping changes
of $(+8.9,-2.94)$ and $(-7.68,+4.62)$ percentage points,
respectively. These opposing directions make simultaneous
improvement of both target modes more difficult than in the
four-bus case.

\begin{figure}[t]
\centering
\includegraphics[scale=0.9]
{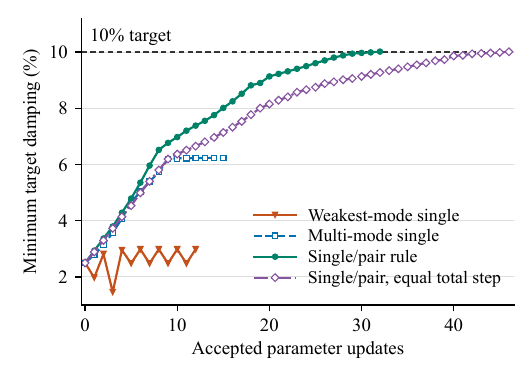}
\caption{IEEE 14-bus minimum target damping versus accepted
updates. The singleton rule stagnates below the $10\%$
requirement, while both single/pair variants reach the target.}
\label{fig:ieee14-damping}
\end{figure}

As shown in Fig.~\ref{fig:ieee14-damping}, the weakest-mode
rule stops after 12 updates, while the multi-mode singleton stops improving after 15 updates with both target damping ratios at
approximately $6.24\%$. In contrast, the proposed single/pair
rule reaches $10.02\%$ and $10.02\%$ in 32 updates.
The equal-total-step variant also reaches the target, requiring
46 updates. These results show that the additional tuning
direction provided by a parameter pair can overcome a local
limitation of singleton tuning even when the total normalized
step is constrained.

Table~\ref{tab:results} summarizes the final damping ratios and
accepted-update counts for both systems. 
In the IEEE 14-bus case, both pairwise tuning methods achieve the required damping level, while the single-parameter tuning methods fail to reach it.

\begin{figure}[t]
\centering
\includegraphics[scale=0.95]
{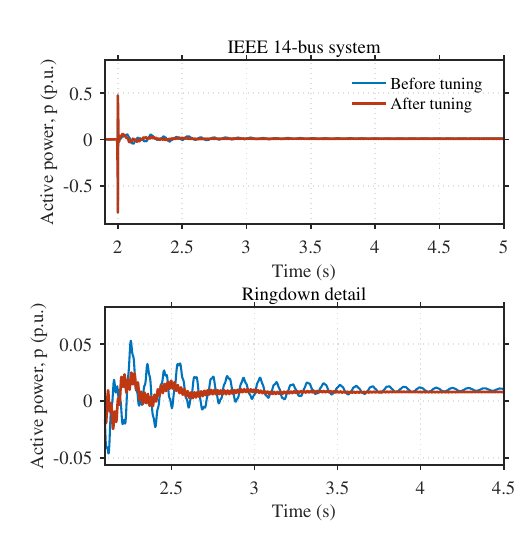}
\caption{IEEE 14-bus EMT active-power response at the bus-3 GFM
converter before and after tuning.}
\label{fig:ieee14-emt}
\end{figure}

The EMT results in Fig.~\ref{fig:ieee14-emt} provide
time-domain confirmation of the improved damping. Although the initial power deviation is similar before and after tuning, but the oscillations decay much faster after tuning. Over $3$--$4$~s, the
peak-to-peak power variation decreases from
$1.9\times10^{-2}$ to $3.5\times10^{-3}$~p.u.,
corresponding to an approximately $81\%$ reduction.
\section{Conclusion}
This paper proposes a conflict-aware impedance-sensitivity-based tuning framework for multi-mode damping enhancement. The method uses normalized damping sensitivities to identify conflicting tuning directions and selects either a single parameter or a coordinated pair through a sparse max-min formulation. Case studies on a four-bus system and a modified IEEE 14-bus system showed that pairwise coordination can reduce tuning effort and overcome cases where singleton tuning stops improving before reaching the damping target. Equal-step comparisons and EMT simulations further confirmed that the improvement comes from coordinated tuning rather than simply larger parameter changes. The impedance-based formulation also avoids requiring complete internal state-space models of all converters.
\section*{Acknowledgment}

This work was supported in part by the U.S. National Science Foundation (NSF) under Award 2552448 and in part  under Award 2509993.

\bibliographystyle{ieeetr}
\bibliography{Ref}

@article{Sun2011,
  author  = {Jian Sun},
  title   = {Impedance-Based Stability Criterion for Grid-Connected Inverters},
  journal = {IEEE Transactions on Power Electronics},
  volume  = {26},
  number  = {11},
  pages   = {3075--3078},
  year    = {2011},
  doi     = {10.1109/TPEL.2011.2136439}
}

@article{Wang2019,
  author  = {Xiongfei Wang and Frede Blaabjerg},
  title   = {Harmonic Stability in Power Electronic-Based Power Systems: Concept, Modeling, and Analysis},
  journal = {IEEE Transactions on Smart Grid},
  volume  = {10},
  number  = {3},
  pages   = {2858--2870},
  year    = {2019},
  doi     = {10.1109/TSG.2018.2812712}
}

@article{Zhu2022Participation,
  author  = {Yue Zhu and Yunjie Gu and Yitong Li and Timothy C. Green},
  title   = {Participation Analysis in Impedance Models: The Grey-Box Approach for Power System Stability},
  journal = {IEEE Transactions on Power Systems},
  volume  = {37},
  number  = {1},
  pages   = {343--353},
  year    = {2022},
  doi     = {10.1109/TPWRS.2021.3088345}
}

@article{Zhu2023Sensitivity,
  author  = {Yue Zhu and Yunjie Gu and Yitong Li and Timothy C. Green},
  title   = {Impedance-Based Root-Cause Analysis: Comparative Study of Impedance Models and Calculation of Eigenvalue Sensitivity},
  journal = {IEEE Transactions on Power Systems},
  volume  = {38},
  number  = {2},
  pages   = {1642--1654},
  year    = {2023},
  doi     = {10.1109/TPWRS.2022.3179143}
}

@article{jankovic2023centralised,
  title={Centralised multimode power oscillation damping controller for photovoltaic plants with communication delay compensation},
  author={Jankovic, Njegos and Rold{\'a}n-P{\'e}rez, Javier and Prodanovic, Milan and Rouco, Luis},
  journal={IEEE Transactions on Energy Conversion},
  volume={39},
  number={1},
  pages={311--321},
  year={2023},
  publisher={IEEE}
}

@article{mateu2025power,
  title={Power oscillation damping controllers for grid-forming power converters in modern power systems},
  author={Mateu-Barriendos, Elia and Alican, Onur and Renedo, Javier and Collados-Rodriguez, Carlos and Almenta, Macarena Martin and Martinez, Edgar Nu{\~n}o and Gomis-Bellmunt, Oriol and Prieto-Araujo, Eduardo},
  journal={IEEE Transactions on Power Systems},
  volume={40},
  number={6},
  pages={5130--5143},
  year={2025},
  publisher={IEEE}
}

@article{feng2025hybrid,
  title={Hybrid oscillation damping and inertia management for distributed energy resources},
  author={Feng, Cheng and Huang, Linbin and He, Xiuqiang and Wang, Yi and D{\"o}rfler, Florian and Kang, Chongqing},
  journal={IEEE Transactions on Power Systems},
  volume={40},
  number={6},
  pages={5041--5056},
  year={2025},
  publisher={IEEE}
}

@article{hoang2026fedavgim,
  title={FedAvgIm: Federated Learning for Multi-Site Impedance Identification of Inverter-Based Resources},
  author={Hoang, Quang Manh and Hollweg, Guilherme Vieira and Hussain, Akhtar and Su, Wencong and Bui, Van-Hai},
  journal={IEEE Transactions on Smart Grid},
  year={2026},
  publisher={IEEE}
}

@article{Cespedes2014,
  author={Cespedes, Mauricio and Sun, Jian},
  title={Impedance Modeling and Analysis of Grid-Connected Voltage-Source Converters},
  journal={IEEE Transactions on Power Electronics},
  volume={29}, number={3}, pages={1254--1261}, year={2014},
  doi={10.1109/TPEL.2013.2262473}
}

@article{Rygg2016,
  author={Rygg, Atle and Molinas, Marta and Zhang, Chen and Cai, Xu},
  title={A Modified Sequence-Domain Impedance Definition and Its Equivalence to the {$dq$}-Domain Impedance Definition for the Stability Analysis of {AC} Power Electronic Systems},
  journal={IEEE Journal of Emerging and Selected Topics in Power Electronics},
  volume={4}, number={4}, pages={1383--1396}, year={2016},
  doi={10.1109/JESTPE.2016.2588733}
}

@article{Harnefors2007,
  author={Harnefors, Lennart and Bongiorno, Massimo and Lundberg, Stefan},
  title={Input-Admittance Calculation and Shaping for Controlled Voltage-Source Converters},
  journal={IEEE Transactions on Industrial Electronics},
  volume={54}, number={6}, pages={3323--3334}, year={2007},
  doi={10.1109/TIE.2007.904022}
}

\end{document}